\PassOptionsToPackage{unicode}{hyperref}
\PassOptionsToPackage{hyphens}{url}
\PassOptionsToPackage{dvipsnames,svgnames,x11names}{xcolor}
\documentclass[12pt]{article}

\usepackage{amsmath,amssymb,amsthm} 
\usepackage[T1]{fontenc}
\usepackage[utf8]{inputenc}
\usepackage{lmodern}
\usepackage{upquote}{}
\usepackage[]{microtype}
\usepackage{parskip}
\usepackage{xcolor}
\makeatletter
\ifx\paragraph\undefined\else
  \let\oldparagraph\paragraph
  \renewcommand{\paragraph}{
    \@ifstar
      \xxxParagraphStar
      \xxxParagraphNoStar
  }
  \newcommand{\xxxParagraphStar}[1]{\oldparagraph*{#1}\mbox{}}
  \newcommand{\xxxParagraphNoStar}[1]{\oldparagraph{#1}\mbox{}}
\fi
\ifx\subparagraph\undefined\else
  \let\oldsubparagraph\subparagraph
  \renewcommand{\subparagraph}{
    \@ifstar
      \xxxSubParagraphStar
      \xxxSubParagraphNoStar
  }
  \newcommand{\xxxSubParagraphStar}[1]{\oldsubparagraph*{#1}\mbox{}}
  \newcommand{\xxxSubParagraphNoStar}[1]{\oldsubparagraph{#1}\mbox{}}
\fi
\makeatother

\usepackage{longtable,booktabs,array}
\usepackage{calc} 
\usepackage{etoolbox}
\makeatletter
\patchcmd\longtable{\par}{\if@noskipsec\mbox{}\fi\par}{}{}
\makeatother
\usepackage{footnotehyper}
\makesavenoteenv{longtable}
\usepackage{graphicx}
\makeatletter
\def\maxwidth{\ifdim\Gin@nat@width>\linewidth\linewidth\else\Gin@nat@width\fi}
\def\maxheight{\ifdim\Gin@nat@height>\textheight\textheight\else\Gin@nat@height\fi}
\makeatother
\setkeys{Gin}{width=\maxwidth,height=\maxheight,keepaspectratio}
\makeatletter
\def\fps@figure{htbp}
\makeatother

\makeatletter
\@ifpackageloaded{caption}{}{\usepackage{caption}}
\AtBeginDocument{%
\ifdefined\contentsname
  \renewcommand*\contentsname{Table of contents}
\else
  \newcommand\contentsname{Table of contents}
\fi
\ifdefined\listfigurename
  \renewcommand*\listfigurename{List of Figures}
\else
  \newcommand\listfigurename{List of Figures}
\fi
\ifdefined\listtablename
  \renewcommand*\listtablename{List of Tables}
\else
  \newcommand\listtablename{List of Tables}
\fi
\ifdefined\figurename
  \renewcommand*\figurename{Figure}
\else
  \newcommand\figurename{Figure}
\fi
\ifdefined\tablename
  \renewcommand*\tablename{Table}
\else
  \newcommand\tablename{Table}
\fi
}
\@ifpackageloaded{float}{}{\usepackage{float}}
\floatstyle{ruled}
\@ifundefined{c@chapter}{\newfloat{codelisting}{h}{lop}}{\newfloat{codelisting}{h}{lop}[chapter]}
\floatname{codelisting}{Listing}

\makeatother
\makeatletter
\@ifpackageloaded{caption}{}{\usepackage{caption}}
\@ifpackageloaded{subcaption}{}{\usepackage{subcaption}}
\makeatother
\usepackage[]{natbib}
\usepackage{bookmark}

\usepackage{xurl} 
\hypersetup{
  pdftitle={Bayesian Fréchet Regression},
  pdfauthor={Simon Fontaine; Bing Li; Lingzhou Xue},
  pdfkeywords={3 to 6 keywords, that do not appear in the title},
  colorlinks=true,
  linkcolor={blue},
  filecolor={Maroon},
  citecolor={Blue},
  urlcolor={Blue},
  pdfcreator={LaTeX via pandoc}}

\newcommand{\indep}{\, \;\, \rule[0em]{.03em}{.65em} \hspace{-.45em}\rule[-.05em]{.65em}{.03em} \hspace{-.45em}\rule[0em]{.03em}{.65em}\;\,\, }

\usepackage{amsmath}
\usepackage{amssymb}
\usepackage{bbm}
\usepackage{mathabx}
\newcommand{\eps}{{\varepsilon}}
\newcommand{\bone}{{\mathbf 1}}

\newcommand{\bbS}{{\mathbb S}}
\newcommand{\bbR}{{\mathbb R}}
\newcommand{\cD}{\mathcal D}
\newcommand{\cF}{\mathcal F}
\DeclareMathOperator{\iid}{iid}
\newcommand{\GP}{\mathcal {GP}}

\DeclareMathOperator{\ran}{ran}

\def\lo{_}
\def\hi{^}

\usepackage[noabbrev, nameinlink]{cleveref}
\crefname{equation}{}{}
\crefformat{equation}{(#2#1#3)}
\crefrangeformat{equation}{(#3#1#4) to~(#5#2#6)}
\crefmultiformat{equation}{(#2#1#3)}%
{ and~(#2#1#3)}{, (#2#1#3)}{ and~(#2#1#3)}

\newcommand{\bbar}{\mathrel{\brokenvert}}
\def\shortbar{\rule[-.0em]{.04em}{.25em}}
\def\bbar{\mbox{\raisebox{-.05in}{$\,\overset{\overset{\shortbar}{\shortbar}}{\shortbar}\,$}}}

\newtheorem{definition}{Definition}
\newtheorem{theorem}{Theorem}
\newtheorem{proposition}[theorem]{Proposition}
\newtheorem{corollary}[theorem]{Corollary}

\newtheorem{assumption}{Assumption}

\usepackage[linesnumbered,ruled,vlined]{algorithm2e}
\usepackage{makecell}

\makeatletter
\newcommand{\dummylabel}[2]{\def\@currentlabel{#2}\label{#1}}
\makeatother

\begin{document}


\title{\bf Bayesian Global Fréchet Regression via Weak Conditional Expectations}
\author{Simon Fontaine, Bing Li, and Lingzhou Xue
\\
Department of Statistics, The Pennsylvania State University}
\date{}
\maketitle

\bigskip
\begin{abstract}
Fréchet regression provides a versatile framework for modeling responses in metric spaces with Euclidean predictors, yet current methodologies rely almost exclusively on frequentist approaches. We propose a Bayesian framework for Fréchet regression that offers a principled way of incorporating prior information into nonlinear global Fr\'echet regression. By targeting a novel Fr\'echet Bayes rule, we reduce the object-valued regression problem to a collection of tractable scalar regression tasks. Our approach allows for a controlled interpolation between the prior and the data-driven frequentist estimate, facilitating effective shrinkage toward informed values. While initially derived under Gaussian assumptions, we demonstrate that our framework is robust to model misspecification by establishing its validity under moment conditions via weak conditional expectations. The numerical properties of the proposed methodology are demonstrated in simulation studies and an application to microbiome compositional data, where we show that leveraging an auxiliary cohort to inform the prior significantly enhances predictive performance in a targeted, small-scale study.
\end{abstract}

\noindent
{\it Keywords:} random objects, metric spaces, nonlinear regression, posterior predictive, weak conditional expectation

\def\spacingset#1{\renewcommand{\baselinestretch}%
{#1}\small\normalsize} \spacingset{1.7}

\spacingset{1.6}

\section{Introduction}\label{sec:introduction}

Regression models aim to characterize the association between a response $Y$ and a set of predictors $X$, generally to facilitate inference or prediction. While a vast literature exists on regression with Euclidean responses—ranging from classical linear and generalized linear models to flexible nonparametric methods such as kernel smoothing, splines, and neural networks—the development of regression methodologies for non-Euclidean responses remains comparatively limited. Such regression problems are becoming increasingly prevalent in modern applications, where the response variable may take the form of a graph, a probability distribution, or an element of a Riemannian manifold. In these settings, the response resides in a metric space that generally lacks the vector-space structure required by conventional regression techniques, rendering many classical methods inapplicable.

The seminal work \cite{petersen_frechet_2019} introduced \textit{global linear Fréchet regression} as a general framework for regression where the object-valued response $Y$ resides in an arbitrary metric space $(\Omega \lo Y,d)$ and predictor $X$ takes values in $\mathbb{R}^p$. This formulation accommodates a diverse range of non-Euclidean data types, including covariance and correlation matrices, functions and distributions, graphs, trees, networks, and points on the sphere.

Building on this foundation, the field has expanded rapidly to include local regression \citep{petersen_frechet_2019,qiu_random_2024}, variable selection and regularization \citep{lin_total_2021,tucker_variable_2023,yang2025variable},
analysis of variance \citep{dubey_frechet_2019},
dimension reduction \citep{ying_frechet_2022,zhang_dimension_2024,zhang_nonlinear_2024}, 
causal inference \citep{lin_causal_2023,bhattacharjee_doubly_2025,tan2026unified}, and others. Notably, \cite{bhattacharjee_nonlinear_2025} proposed a global nonlinear Fréchet regression framework for object-valued response with object-valued predictors through reproducing kernel Hilbert space (RKHS) embeddings and the notion of a weak conditional Fréchet mean (see their Section 3 for the details), where the global linear Fréchet regression emerges as a special case by choosing a linear kernel. 

While these frequentist developments have expanded the flexibility of Fréchet models, they leave a gap in settings where the integration of prior knowledge is important. The Bayesian paradigm provides a principled mechanism to fill this gap, enabling the incorporation of external or prior knowledge to guide or shrink estimates towards informed values. This is particularly relevant in small to modest sample regimes where frequentist estimates may exhibit high variance or fail to generalize. Specifically, leveraging a larger, auxiliary dataset to construct an informative prior can yield substantial gains in predictive accuracy and inferential stability. This situation often occurs when data from a broad population study can be used to inform the analysis of a more targeted, resource-intensive sub-population. We illustrate this utility in Section~\ref{sec:microbiome} through an application to microbiome compositional data, demonstrating that regression performance in a specialized cohort can be significantly enhanced by shrinking toward estimates derived from a larger, more general population.

The Bayes rule occupies a central role in Bayesian inference, as it minimizes the posterior expected loss and is therefore optimal from a decision-theoretic perspective. To formulate a Bayes rule for Fr\'echet regression is goal of this paper. 
However, unlike in classical regression where the response resides in a vector space, formulating a Bayes rule for a metric-space-valued response is by no means straightforward. The primary challenge is that there is no linear operation or integral in a generic metric space, which means direct construction of prior distribution,  likelihood,  or posterior distribution on it is all but impossible, unless additional structures are imposed on the metric space.  
While some recent research has explored this direction with a Bayesian flavor, such as Stein shrinkage estimation for the Fréchet mean \citep{mccormack_stein_2022}, there is currently no general framework for incorporating specific prior knowledge into a Fréchet regression model.

\def\real{\mathbb{R}}

We address this fundamental difficulty by bypassing the construction of a prior distribution and the likelihood on the metric space, and shifting attention to the metric itself, which is a real-valued random variable that permits the construction of a prior and a likelihood. The focus on metric is justified because essentially all properties of a metric space are derived from the properties of the metric itself (see, for example, \cite{kelley1955general}, Chapter 4), and in this sense, a metric space is nothing more than the metric itself.   Specifically, suppose $Y \lo x$ is a metric-space-valued response observed at predictor $x$, we assume $E(d \hi 2 (Y \lo x, y) |f) = f(x)$, where $f$ is a random element in a reproducing kernel Hilbert space (RKHS) with a prior distribution. We then construct a likelihood based on the sample of squared metrics: 
\begin{align*}
    \{ d \hi 2 (Y \lo 1, y), \ldots, d \hi 2 (Y \lo n, y) \},  
\end{align*}
and introduce a prior distribution for $f$ in the RKHS, both of which are feasible as we are dealing with vector spaces. 
The likelihood on $\real$ and the prior distribution on the RKHS allow us to find the Bayes rule for $E(d \hi 2 (Y \lo x, y) | f)$, which is then minimized over $y$ to obtain our regression estimate. This estimate is defined as the Bayes rule for Fr\'echet regression, or simply the Fr\'echet Bayes rule, the central theme of our paper. In addition, the Fréchet Bayes rule is not restricted to the current context of Fréchet regression and can potentially provide a general mechanism for Bayesian inference in metric spaces, which deserve further investigation.

This construction bypasses the need for constructing a full generative model from inside the metric space. The above Bayesian formulation with an RKHS regression function fits naturally within the global Fr\'echet regression paradigm of \cite{bhattacharjee_nonlinear_2025}. Indeed, when the sample size $n \to \infty$, the above Bayesian procedure reduces to the global Fr\'echet regression. When the sample size is modest, it offers a principled way of incorporating prior information about $f$ into the global Fr\'echet regression. 

We first develop this approach under Gaussian assumptions to derive the Fr\'echet Bayes rule. We subsequently relax the distributional assumption to weaker moment conditions in the context of weak conditional expectations \citep{li_dimension_2022,bhattacharjee_nonlinear_2025,sang2026nonlinear}. 

We summarize the primary contributions of this work as threefold. 
\begin{itemize}
    \item We introduce a principled Bayesian framework for Fréchet regression that bypasses the need for a generative likelihood. To the best of our knowledge, this represents the first systematic Bayesian treatment of Fréchet regression in the literature. By targeting the novel Bayes rule for $E( d \hi 2 (Y \lo x, y) |f)$ based on $\{d \hi 2 (Y\lo i, y): i=1, \ldots, n\}$, we reduce the complex, object-valued regression problem into a collection of tractable scalar regression tasks. This serves as a foundational bridge towards Bayesian analysis in arbitrary metric spaces. 
    \item Our approach enables the direct incorporation of explicit prior knowledge to guide the Fréchet regression function. This represents a significant advancement over existing methods to allow for true posterior shrinkage toward informed, external estimates. Furthermore, this framework provides a general mechanism for Bayesian prior elicitation in random object analysis where a generative likelihood is unavailable.
    \item We establish the solid theoretical foundations for our procedure: a Gaussian framework where posterior updates are exact, and a robust alternative based on weak conditional expectations that requires only weaker moment conditions, ensuring the inferential robustness of our procedure under minimal distributional assumptions.
\end{itemize}

Through both simulated and real-world data examples, we demonstrate that our proposed Bayesian Fréchet regression provides a principled interpolation between the frequentist estimate and the prior. Importantly, our results show that even under misspecified priors, the Bayesian Fréchet regression offers substantial gains in predictive accuracy and stability.

The rest of this paper is organized as follows. 
Section~\ref{sec:prelim} provides the preliminaries. Section \ref{subsection:from B to FB} introduces the notion of Fr\'echet Bayes rule, and Section~\ref{sec:bfr} develops the Bayesian global Fréchet regression under a working Gaussian model. Section~\ref{sec:fbr} develops the weak Fréchet Bayes rule for global Fréchet regression to relax distributional assumptions using weak conditional expectations. Section \ref{sec:implement} presents the prior specification and implementation strategies. We demonstrate the empirical performance of our framework in Sections~\ref{sec:sims} and \ref{sec:microbiome} through simulation studies and an application to microbiome compositional data. Section \ref{sec:discussion} includes a few concluding remarks. Proofs of the main results, additional technical details, and numerical results are provided in the Supplementary Materials.

\section{Preliminaries}\label{sec:prelim}


\def\ca#1{{\cal {#1}}}

\textbf{Metric spaces and random objects.} 
A \textit{metric space} consists of a set $\Omega \lo Y$ equipped with a \textit{metric} (or distance function) $d :\Omega \lo Y\times\Omega \lo Y\to [0, \infty)$ that satisfies the standard axioms of positive definiteness, symmetry, and the triangle inequality. Let $(\Omega, \cF, P)$ denote an underlying probability space, where $\Omega$ is the sample space, $\cF$ is a $\sigma$-algebra over $\Omega$, and $P$ is a probability measure over $\cF$. Let $\ca F \lo Y$ be the Borel $\sigma$-algebra generated by open sets in $\Omega \lo Y$. A random element $Y:\Omega\to\Omega \lo Y$ that is measurable with respect to $\cF/\ca F \lo Y$ is called a  \textit{random object},  to distinguish it from random vectors in an Euclidean space or random functions in  Hilbert spaces, which is important when $\Omega \lo Y$ lacks a vector-space structure. The distribution of $Y$,  written as $P \lo Y$, is the   push-forward measure $P_Y=P\circ Y^{-1}$ on $(\Omega \lo Y, \ca F \lo Y)$. 

\def\var{\mathrm{var}}
\def\cov{\mathrm{cov}}
\def\tr{\mathrm{tr}}

In a general metric space, in contrast to Euclidean or Hilbert spaces, the absence of vector-space operations necessitates a reliance on the metric structure for computation, comparison, and the characterization of centrality. \cite{frechet_elements_1948} proposed the \textit{Fréchet mean} (also known as \textit{barycenter}) as a measure of centrality for a random object. It is defined as the set of minimizers of the expected squared distance:
$$    E_{\oplus} (Y) = \underset{y\in\Omega \lo Y}{\arg\min}\; E \{d ^2(Y, y)\} \subset\Omega \lo Y.$$
In Euclidean spaces, the Fréchet mean coincides with the classical mean. The corresponding minimum value is the \textit{Fréchet variance}, which quantifies the dispersion of $Y$ around its Fréchet mean:
   $ \var_{\oplus} (Y) = \underset{y\in\Omega \lo Y}{\min}\; E \{d ^2(Y, y )\}.$
 In the Euclidean setting, the Fréchet variance reduces to the \textit{total variance}, $\tr[\cov(Y)]$, rather than the covariance matrix $\cov(Y)$, reflecting the aggregate variability across all dimensions.

\textbf{Global Fréchet regression.}
\cite{petersen_frechet_2019} introduced the (linear) global \textit{Fréchet regression} for metric-space-valued responses $Y\in\Omega \lo Y$ and Euclidean predictor $x\in\bbR^p$.
The core insight comes from the observation that, in the classical linear regression with Gaussian error,  the conditional expectation in Euclidean space can be characterized entirely through the minimization of an objective function of the form $E[ s(X,x) ( Y - y ) \hi 2 ]$, where $s(X,x)$ is a linear function of $X$ with expectation 0. 
  By replacing the Euclidean metric with a general distance $d$ over the response space $\Omega \lo Y$, as in Equation (2.8) in Section 2.2 of \cite{petersen_frechet_2019}, the global linear Fréchet regression 
takes the form of
\begin{align}
\underset{y\in\Omega \lo Y}{\arg\min}\; E \{s(X, x)d ^2(Y, y)\}.
    \label{eq:bfr.gfr.objective}
\end{align}
where $s(X,x)$ is the same weight as appeared in 
classical linear regression.  

\def\real{\mathbb{R}}
\def\trans{^\top}
\def\argmin{\mathrm{argmin}}

 Motivated by this insight, \cite{bhattacharjee_nonlinear_2025} recast the objective function in (\ref{eq:bfr.gfr.objective}) within the framework of \textit{weak conditional expectations} proposed by \cite{li_dimension_2022}; see also \cite{song-dimension-2017}. Indeed, the objective function in (\ref{eq:bfr.gfr.objective}) is exactly the $L \lo 2 (P)$-projection of $d \hi 2 (Y, y)$ on to the linear space $\{a + b\trans X: b \in \real, b \in \real \hi p \}$, which is a special form of the weak conditional expectation developed in \cite{li_dimension_2022}.  \cite{bhattacharjee_nonlinear_2025} augmented the space of linear functions of $X$ to an RKHS of functions of $f$ (plus a constant term if the RKHS doesn't contain 1), resulting in the {\em nonlinear global Fr\'echet regression}. In terms of the weak conditional expectation relative to an RKHS, the nonlinear global Fr\'echet regression estimator takes the form 
\begin{align*}
E \lo \oplus (Y \bbar X) =   \underset{y \in \Omega \lo Y}  \argmin \,  E [ d \hi 2 (Y, y) \bbar X], 
\end{align*}
where $E [ d \hi 2 (Y, y) \bbar X]$ is the weak conditional expectation of $d \hi 2 (Y, y)$. \cite{bhattacharjee_nonlinear_2025} referred to $E \lo \oplus (Y \bbar X)$ as the weak Fr\'echet conditional mean of $Y$ given $X$, as its form mimics the Fr\'echet conditional mean defined as $\argmin \lo {y \in \Omega \lo Y} \, E[ d \hi 2 (Y, y)|X ]$.

\section{From Bayes Rule to Fr\'echet Bayes Rule}\label{subsection:from B to FB}

We first introduce the notion of Fr\'echet Bayes rule, which is constructed by following the underlying logic of the classical Bayes rule and Fr\'echet conditional mean. Suppose that $Y \lo 1, \ldots, Y \lo n$ are a sample of random variables or random vectors of size $n$, that $\theta$ is a parameter that determines the distribution of $(Y \lo 1, \ldots, Y \lo n)$, and that $\theta$ is a random variable with a prior density $\pi (\theta)$, supported on a parameter space. Suppose that we are interested in estimating a parameter $t (\theta)$ that lives in the same space as $Y$. The Bayes rule estimating $t (\theta)$ with respect to the loss function $L(Y, a) = \| Y - a \| \hi 2$ is simply the posterior mean 
\begin{align*}
    E ( t(\theta) | Y \lo 1, \ldots, Y \lo n ). 
\end{align*}

To facilitate the subsequent development, we now extend the concept of Bayes rule to the case where $t(\theta), Y \lo 1, \ldots, Y \lo n$  are random objects taking values in a metric space. The primary challenge in this extension is that the counterpart of  $E(t (\theta) | Y \lo 1, \ldots, Y \lo n)$ is not defined, because there is neither a linear operation nor an integral in a general metric space.   This absence of a generative structure makes the standard path for prior specification and posterior updating non-obvious.

Our basic idea is to transfer the operation of 
  finding the  Bayes rule for $t(\theta)$ 
from the random-object level to the squared-metric level.   In other words, we first look for the Bayes rule for estimating the parameter $E[ d \hi 2 (Y, y)|\theta ]$ based on the sample $\{ d \hi 2 (Y \lo i, y): i = 1, \ldots, n\}$, and then predict $Y \lo x$ by minimizing this Bayes rule. In fact, the same spirit underlies  the construction of Fr\'echet expectation and conditional Fr\'echet expectation, which can be characterized by two steps: 
\begin{itemize}
    \item Step 1: perform  an operation  (e.g., taking mean or conditional mean) on $d \hi 2 (Y,y)$; \vspace{-.08in}
    \item Step 2: obtain the Fr\'echet counterpart of this operation  (e.g., Fr\'echet mean and conditional Fr\'echet mean of $Y$) by minimizing the objective function obtained from Step 1 over $y \in \Omega \lo Y$. 
\end{itemize} 
Thus, in the present Bayesian-Fr\'echet context, we simply substitute the ``operation'' in the above general scheme by ``finding the Bayes rule of'',  and substitute the target $  d \hi 2 ( Y, y)$ by $E[d \hi 2 (Y, y)|\theta]$. We now formally define the Fr\'echet Bayes rule as follows. 

\begin{definition}\label{definition:frechet bayes rule} The Fr\'echet Bayes (FB) rule for estimating $E \lo \oplus (Y|\theta)$ is 
\begin{align*}
\hat Y \lo {\mathrm{FB}} =    \underset{y \in \Omega \lo Y}{\argmin} \, E [E( d \hi 2 (Y, y)|\theta )| d \hi 2 (Y \lo 1, y), \ldots, d \hi 2 (Y \lo n, y)]. 
\end{align*}
\end{definition}
In other words, we are treating the objective function for a fixed $y$ as our Bayesian estimand, and then minimizing the Bayes rule over $y$. 
By the iterative law of conditional expectation,  the Fr\'echet Bayes rule can be expressed as 
\begin{align*}
    \underset{y \in \Omega \lo Y}{\argmin} \, E [  d \hi 2 (Y, y)  | d \hi 2 (Y \lo 1, y), \ldots, d \hi 2 (Y \lo n, y)]. 
\end{align*}

\section{Bayesian Global Fréchet Regression}\label{sec:bfr}

Our goal is to propose a Bayesian framework of nonlinear global Fr\'echet regression that allows us to incorporate prior information about the regression function. We first review the Bayes rule for classical regression in Section \ref{subsec:classical} and then introduce the Bayes rule for Fr\'echet regression in Section \ref{subsec:bfr.fbr}. In Section \ref{subsec:bfr.nlbr}, we give a full construction of the Bayesian global Fr\'echet regression model under a working Gaussian model.


\def\lo{_}
\def\real{\mathbb{R}}
\def\trans{^\top}
\def\hi{^}

\subsection{Bayes rule for classical regression}\label{subsec:classical}

Suppose $(x \lo 1, Y \lo 1), \ldots, (x \lo n, Y \lo n)$ are observed samples of predictors and responses, where $x \lo i \in \real \hi p$ are nonrandom vectors, and $Y \lo i \in \real$ are random responses. Suppose  $(x,Y \lo x)$ is a new pair of predictor and response, where $x$ is observed and $Y \lo x$ is unobserved and is to be predicted. The general goal of regression is to estimate $E(Y \lo x )$ based on $(x \lo 1, Y \lo 1), \ldots, (x \lo n, Y \lo n)$.   Usually, $E(Y \lo x )$ is written as $E(Y|x)$, but since we assume $x$ is nonrandom, it is more rigorous to write $E(Y \lo x)$. Treating $X$ as random (as suggested by the notation $E(Y|x)$) would be especially confusing since later on our prior distribution will involve  $x$. The notation $Y \lo x$ implies that $Y$ is a random function of $x$, which will be made rigorous as we proceed further. 
For example, in linear regression, we assume  $Y \lo x = \beta \trans x + \epsilon$, where $Y \lo x$ is indeed a random function of $x$ (meaning the distribution of $Y \lo x$ depend on $x$). In this notation our goal is then to estimate $E (Y \lo x) = \beta \trans x$,  or equivalently $\beta$.  

In the Bayesian linear regression setting  (see, for example, \cite{bishop_pattern_2006} and \cite{gelman_bayesian_2021}), we assign a prior density $\pi (\beta)$ on $\beta$, and aim to estimate the transformed parameter $t \lo x(\beta)= \beta \trans x$. The Bayes rule for this transformed parameter under the squared loss is posterior mean  $E(t \lo x (\beta)|  Y \lo 1, \ldots, Y \lo n)$. Again, in the Bayesian regression literature, this is often written as $E(t \lo x(\beta) |x \lo 1, Y \lo 1, \ldots, x \lo n Y \lo n)$, but we omit $x \lo 1, \ldots, x \lo n$ in the conditioned argument for the same reason mentioned earlier. More generally, for general nonlinear regression,  we assume $E(Y \lo x|f) = f(x)$, where $f$ is an unknown random function, and put a prior $\pi (f)$ on  $f$. The Bayes rule for regression under the squared loss is then the posterior mean 
$ 
 E[f(x) |  Y \lo 1, \ldots,   Y \lo n].   
$
This is called {\em Bayes rule for regression}. By the Iterative law of conditional expectation, the above can be written as 
\begin{align*}
    E [ E (Y \lo x  | f) | Y \lo 1, \ldots, Y \lo n] = E (Y \lo x  | Y\lo 1, \ldots, Y \lo n). 
\end{align*}
The right-hand side is commonly known as the {\em posterior predictive mean}. 
\subsection{Bayes rule for Fréchet regression}\label{subsec:bfr.fbr}

\def\ca#1{{\cal{#1}}}

We now introduce the Bayesian nonlinear global Fréchet regression.   
Let  $\Omega \lo X$ be a subset of a Hilbert space representing the predictor space. 
Echoing the $Y \lo x$ notation in the last subsection, we will, without loss of generality, assume that the responses are evaluations of a random function of  $x$. Let $F \lo d \hi 2 (P)$ be the collection of random elements $W$ in $(\Omega \lo Y , d)$ such that $E ( d \hi 2 (W, y)) < \infty$ for all $y \in \Omega \lo Y $; that is
\begin{align*}
    F \lo d \hi 2 (P) = \left\{ (W: \Omega \to \Omega \lo Y ): \int \lo {\Omega \lo Y } d \hi 2 (W, y) d P < \infty \  \mbox{for all $y \in \Omega \lo Y$} \right\}
\end{align*}
We assume $Y$ to be a mapping from $\Omega \lo X$ to $F \lo d \hi 2 (P)$. This means $Y$ is a random function of $x$ and that, for each $x \in \Omega \lo X$ and each $y \in \Omega \lo Y$, $E[ d \hi 2 (Y \lo x, y) ]< \infty$.  
Let $x \lo 1, \ldots, x \lo 1, x$ be members of $\Omega \lo X$, and let $Y \lo {x \lo 1}, \ldots, Y \lo {x \lo n}, Y \lo x$ be $\Omega \lo Y$-valued responses associated with $x \lo 1, \ldots, x \lo n, x$. For convenience, we will abbreviate $Y \lo {x \lo i}$ as $Y \lo i$, but reserve   $Y \lo x$ for $Y$ evaluated at $x$.

\def\argmin{\mathrm{argmin}}

We now give a formal definition of  {\em Bayes rule for Fr\'echet regression}. 

\begin{definition}
    Suppose $Y \lo x$  is a member of $F \lo d \hi 2 (P)$. The Fr\'echet Bayes rule for regression is 
\begin{align}\label{eq:objective function for Bayes rule}
   \underset{y \in {\Omega \lo Y }}{\argmin}  \,   E [d  \hi 2  ( Y \lo x, y)   | d \hi 2 (Y \lo 1,y), \ldots, d \hi 2 (Y \lo n,y) ].
\end{align}    
\end{definition}

\medskip

For convenience, throughout the rest of the paper, we use $a \lo {1:n}$ to represent a vector $(a \lo 1, \ldots, a \lo n)\trans$. In addition, we will use the following abbreviations: 
 \begin{align*}
     U \lo i (y) = d \hi 2 (Y \lo i, y), \quad U \lo x (y) = d \hi 2 (Y \lo x, y). 
 \end{align*}

\def\ka{\kappa}
\def\inv{^{-1}}

\subsection{Bayesian Fréchet regression under a working Gaussian model}\label{subsec:bfr.nlbr}

The computation of the  {Fr\'echet Bayes rule} requires access to the conditional expectation $  E  [   U \lo x (y) | U \lo {1:n}(y)]$. 
For a fixed pair $(x,y)\in \Omega \lo X \times \Omega \lo Y$, this task reduces to a  {Bayesian} scalar regression problem with a real-valued response $d \hi 2 (Y \lo x, y) \in\bbR$,  a predictor  $x\in{\Omega \lo X} $  {and a random function $f$ as the random parameter.}

\def\var{\mathrm{var}}

\def\ali{\,&}

In this section we assume $f$ to be a Gaussian random element in an RKHS $\ca H $ generated by a positive kernel $\ka: \Omega \lo X \times \Omega \lo X \to \real$ and spanned by the set of functions $\{\ka (\cdot,x \lo 1), \ldots, \ka (\cdot, x \lo n)\}$. 
We regard this model as a working model because $d^2$ is positive and cannot have a Gaussian distribution in the proper sense.
A more rigorous development will be carried out in Section~\ref{sec:fbr}, which removes this rigid Gaussian assumption.
Some kernels, such as the commonly used  Gaussian radial basis kernel, do not contain nonzero constant functions. For this reason, we will separate the intercept from the nonlinear regression function in the RKHS.  Specifically, let $\mu \in \ca H$ be the function $n\inv \sum \lo {i=1} \hi n \ka (\cdot, x \lo i)$ and let $\ka \lo c(\cdot, x)$ be the centered kernel $\ka (\cdot, x) - \mu$. Resetting $\ca H$ to be the centered RKHS spanned by $\{\ka \lo c (\cdot, x \lo 1), \ldots, \ka \lo c (\cdot, x \lo n ) \}$, we make the following assumption. 
\begin{assumption}[Bayesian Fr\'echet regression model with Gaussian likelihood and prior]\label{assumption:Bayesian Frechet Gaussian prior} \ 
\begin{enumerate}
\item  For any $x \in \Omega \lo X$,   
$
U \lo x (y)|c,f    \sim N(  c  +  \langle  f , \ka \lo c  (\cdot, x ) \rangle \lo {\ca H} , \sigma \lo y \hi 2 )$,
where $c$ is a random variable in $\real$ and $f$ is a random function in $\ca H$;   \vspace{-.07in}
\item $Y \lo 1 \indep \cdots \indep Y \lo n \indep Y \lo x | (f,c)$; \vspace{-.07in}
\item    $ c   \sim   N( \alpha \lo y, \tau \lo y \hi 2 ) \ \mbox{for some $\alpha \lo y \in \real$,  $\tau \lo y \hi 2 > 0$ }, $  \vspace{-.07in}
\item    $ f   \sim   N(\nu \lo y, \Lambda  \lo y )$,    $\nu \lo y \in \ca H$, $\Lambda \lo y: \ca H\to \ca H$ is a linear operator, and $  c    \indep f $. 
\end{enumerate}
\end{assumption}

\newcommand{\barcHX}{{\overline{\cH}_{\Omega \lo X} }}
\newcommand{\barcB}{{\overline{\cB}}}

The hyperparameters $\alpha \lo y, \tau \lo y\hi 2, \sigma \lo y \hi 2 , \nu \lo y$,  and $\Lambda \lo y$   are elicited specifically for each $y$ as detailed in Section~\ref{subsec:bfr.prior}. Note that part 1 of Assumption \ref{assumption:Bayesian Frechet Gaussian prior} can be alternatively written in regression form as 
\begin{align*}
    U \lo x(y) = c + f(x) - n\inv \sum \lo {i=1} \hi n f(x \lo i) + \epsilon, 
\end{align*}
where $E(\epsilon | c, f) = 0$,  $\var(\epsilon | c, f) = \sigma \lo y \hi 2$.

\def\cran{\overline{\mathrm{ran}}}
\def\ran{\mathrm{ran}}
\def\cran{\overline{\ran}}

The conjugacy of the model in Assumption \ref{assumption:Bayesian Frechet Gaussian prior} ensures that the joint posterior for $(c, f)$ is still Gaussian. In the following, let $B: \ca H \to \real \hi n$ be the linear operator
\begin{align*}
    B f = ( \langle f, \ka \lo c (\cdot, x \lo 1) \rangle \lo {\ca H}, \ldots, \langle f, \ka \lo c (\cdot, x \lo n) \rangle \lo {\ca H} )\trans. 
\end{align*}
It is easy to verify that the adjoint of this operator is 
$B \hi * a = \sum \lo {i=1} \hi n a \lo i \ka \lo c (\cdot, a \lo i), $
where $a = (a \lo 1, \ldots, a \lo n)$ is an arbitrary vector in $\real \hi n$. We will use $\bone \lo n$ to denote $n$-dimensional vector $(1, \ldots, 1) \trans$. For a linear operator $A: \ca H \to \ca A$, let $\ran(A)$ denote the range of $A$ and $\cran(A)$ the closure of $\ran(A)$. For a self-adjoint operator $A$, its restriction on $\cran(A)$  is an injective mapping from $\cran (A)$ to $\ran(A)$. The inverse of this mapping (from $\ran (A)$ to $\cran(A)$)  is called the Moore-Penrose generalized inverse of $A$, denoted by $A \hi \dagger$. 
The next theorem gives the explicit forms of their posterior means of $c$ and $f$. 
\begin{theorem}\label{theorem:posterior means for c and f}
Under Assumption \ref{assumption:Bayesian Frechet Gaussian prior}, the posterior means of $c$ and $f$ are given by
\begin{align}\label{eq:bfr.nlbr.ppm_cm}
\begin{split}
E  [ c | U \lo {1:n}(y)] 
=\ali 
\frac{\alpha \lo y /\tau \lo y \hi 2 + \bone \lo n \trans U \lo {1:n} (y) /\sigma \lo y \hi 2 }{1/\tau \lo y \hi 2 + n/\sigma \lo y \hi 2 }, \\
E  [ f | U \lo {1:n}(y)] 
=\ali 
(B \hi * B/\sigma \lo y \hi 2  + \Lambda \lo y^\dagger)^\dagger [B \hi * U \lo {1:n}(y) /\sigma \lo y \hi 2 + \Lambda \lo y ^\dagger \nu \lo y ].
\end{split}    
\end{align}
\end{theorem}

\medskip 

It immediately follows that the posterior predictive mean $E [U \lo x (y) |Y \lo {1:n}]$ has an analytical form given by the next corollary. 
\begin{corollary}\label{corollary:E U x y}
Under Assumption \ref{assumption:Bayesian Frechet Gaussian prior}, we have 
\begin{align}
  E [U \lo x (y) |U \lo {1:n}(y)]
    &=
     E  [ c | U \lo {1:n}(y)] 
    +
    \langle \ka \lo c (\cdot, x), E  [ f | U \lo {1:n}(y)] \rangle_{\ca H}, \label{eq:bfr.nlbr.ppm_rkhs}
\end{align}
where    $  E  [ c | U \lo {1:n}(y)] $ and 
 $E  [ f | U \lo {1:n}(y)] $ are given in (\ref{eq:bfr.nlbr.ppm_cm}). 
\end{corollary}

The estimator \eqref{eq:bfr.nlbr.ppm_rkhs} provides a principled combination between prior knowledge and data-driven evidence. In the large-sample limit  ($n\to\infty$) or under a vanishing prior ($\tau \lo y \hi 2,\Lambda \lo y\to\infty$), the posterior predictive mean function $E  [U \lo x (y)|Y \lo {1:n}]$ reduces to 
\begin{align*}
    E \lo n [U \lo x (y)] + \langle \ka \lo c (\cdot, x),( B \hi * B ) \hi \dagger  B \hi * U \lo {1:n} (y) \rangle \lo {\ca H}, 
\end{align*}
which is the sample-level objective function for the (frequentist) nonlinear global Fr\'echet regression introduced by \cite{bhattacharjee_nonlinear_2025}, where 
\begin{align*}
    n \inv B \hi * B = n \inv \sum \lo {i=1} \hi n \ka \lo c (\cdot, x \lo i) \otimes \ka \lo c (\cdot, x \lo i)
\end{align*}
is denoted by $\Sigma \lo {XX}$. This operator can be interpreted as the sample covariance operator based on $x \lo 1, \ldots, x \lo n$, although in the present paper, we assume $x \lo 1, \ldots, x \lo n$ to be nonrandom vectors. 
Conversely, as the prior concentrates ($\tau \lo y \hi 2,\Lambda \lo y \hi 2\to 0$), the posterior predictive mean function $E  [ U \lo x (y)|Y \lo {1:n}]$ tends to the prior mean function $\alpha \lo y+\langle \ka \lo c (\cdot, x), \nu \lo y\rangle \lo {\ca H}$, which is completely determined by the prior information. By the continuity of the Fréchet minimizer, these properties extend to the posterior predictive Fréchet mean, which effectively shrinks the frequentist regression function toward the prior Fréchet mean.

While the posterior predictive mean \cref{eq:bfr.nlbr.ppm_rkhs} is defined abstractly in $\ca H$, practical evaluation requires a coordinate representation to transform the Hilbert space inner products into tractable matrix operations. We detail a computationally efficient implementation based on a judicious choice of prior in Section~\ref{sec:implement}.

\section{Weak Fréchet Bayes Rule}\label{sec:fbr}

While the Fréchet Bayes rule introduced in Section~\ref{subsec:bfr.fbr} provides a principled objective, the closed-form expression for the posterior predictive mean in \eqref{eq:bfr.prior.ppm1} was derived under the Gaussian prior assumption specified by Assumption 2.  However, the Gaussian assumption on the squared distances as computed in \cref{eq:bfr.prior.ppm1}, which are inherently non-negative, is primarily a computational convenience. In this section, we remove the rigid Gaussian distribution assumption by replacing the posterior means of $c$ and $f$  with {\em weak posterior means} based on the notion of weak conditional expectation introduced by \cite{li_dimension_2022}. This construction echoes that of nonlinear global Fr\'echet regression \citep{bhattacharjee_nonlinear_2025} in the frequentist setting, which is also based on the weak conditional expectation.

\subsection{Weak conditional expectation}\label{subsec:fbr.lce}

\def\cov{\mathrm{cov}}
\def\hii#1{^{(#1)}}

Weak conditional expectation was introduced in the context of functional sufficient dimension reduction, where both the predictor and the response are Hilbert space-valued random elements. In the following, we gather together some information scattered in \cite{li_dimension_2022}, \cite{li-2018-linear},  \cite{li_sufficient_2018}, and \cite{bhattacharjee_nonlinear_2025} that is relevant to our development. To avoid confusion with the current context,  we use $U$ to represent the predictor and $V$ the response. Suppose, then, that  $U$ and $V$ are random elements taking values in Hilbert spaces $\ca G$ and $\ca H$, respectively, and  $\mathfrak M$ is a Hilbert space consisting of functions defined on $\ca G$ and taking values in $\real$. The definition of weak conditional expectation relies on the following assumption. 
\begin{assumption}\label{assumption:bounded bilinear forms} The following bilinear forms are bounded: 
\begin{align*}
    b \lo 1 : \ali  \mathfrak M \times \mathfrak M \to \real, \quad (f \lo 1, f \lo 2) \mapsto E [f \lo 1(U)f \lo 2(U)], \\
    b \lo 2 : \ali \mathfrak M \times \ca H \to \real, \quad (f, h ) \mapsto E[f(U) \langle V, h \rangle \lo {\ca H}], \\
    b \lo 3: \ali \ca H \times \ca H \to \real, \quad (h \lo 1, h \lo 2) \mapsto E [\langle h \lo 1, U \rangle \lo {\ca H} \langle h \lo 2, U \rangle \lo {\ca H}]. 
\end{align*}
\end{assumption}
This assumption guarantees the unique existence of five quantities on which weak conditional expectation is based.  
\begin{enumerate}
    \item There exist unique   $\mu \lo U \in \mathfrak M $ and $\mu \lo V \in \ca H$  such that,    for all $m \in \mathfrak M$, $h \in \ca H$,  
    \begin{align*}
       E   m(U) = \langle m, \mu \lo U \rangle \lo {\mathfrak M}, \quad E  \langle h, V \rangle \lo {\ca H} = \langle f, \mu \lo V \rangle \lo {\ca H};
    \end{align*}
  \vspace{-.4in}
    \item There exist unique bounded linear operators $\Sigma \lo {UU}: \mathfrak M \to \mathfrak M$, $\Sigma \lo {VV} : \ca H \to \ca H$ and $\Sigma \lo {UV}: \ca H \to \mathfrak M$ such that, 
    for all $f \lo 1, f \lo 2, f \in \mathfrak M$ and all $h \lo 1, h \lo 2, h \in \ca H$,
    \begin{align*}
        \langle f \lo 1, \Sigma \lo {UU} f \lo 2 \rangle \lo {\mathfrak M} = \ali  \cov[f \lo 1 (U), f \lo 2 (U)], \\ \langle h \lo 1, \Sigma \lo {VV} h \lo 2 \rangle \lo {\ca H} = \ali \cov[ \langle h \lo 1, V \rangle \lo  {\ca H}, \langle h \lo 2, V \rangle \lo {\ca H}  ], \\
     \langle f, \Sigma \lo {UV} h \rangle \lo {\mathfrak M} = \ali \cov [  f(U), \langle h , V \rangle \lo {\ca H} ].  
    \end{align*}
    \end{enumerate}
 These unique existence results are derived from the Riesz representation theorem (see, for example, \cite{conway_course_1990}). The Hilbertian elements $\mu \lo U$ and $\mu \lo V$ are called the mean elements of $U$ and $V$, and the linear operators $\Sigma \lo {UU}, \Sigma \lo {UV}, \Sigma \lo {VV}$ are called covariance operators.    
Assuming $\ran ( \Sigma \lo {UV}) \subseteq \ran ( \Sigma \lo {UU})$,   the operator $R \lo {UV} \hii c = \Sigma \lo {UU} \hi \dagger \Sigma \lo {UV}$ is well defined, where $\Sigma \lo {UU} \hi \dagger$ is the Moore-Penrose inverse of $\Sigma \lo {UU}$. This operator is called the {\em centered regression operator}. 

Next, for each $u \in \ca G$, consider the mapping
\begin{align*}
    R \lo {UV} \hii c (\cdot)(u): \, \ca H \to \mathfrak M, \quad h \mapsto (R \lo {UV} \hii c  ( h)) (u). 
\end{align*}
It is easy to see that $R \lo {UV} \hii c (\cdot)(u)$ is a linear functional. We assume that it is a bounded linear functional for every $u \in \ca G$; a linear operator that satisfies this condition is the Carleman operator (see \cite{weidmann1980linear}). We summarize the above conditions in the following assumption. 
\begin{assumption}\label{assumption:carleman}
 \quad   $\ran (\Sigma \lo {UV}) \subseteq \ran ( \Sigma \lo {UU})$  and $R \lo {UV} \hii c$ is a  Carleman operator. 
\end{assumption}
Assumption \ref{assumption:carleman} guarantees the existence of the Riesz representation $\lambda \lo {R \lo {UV} \hii c}(u)$ of the bounded linear functional $R \lo {UV} \hii c (\cdot)(u)$ for each $u \in \ca G$, which leads to our definition of the (centered version of) weak conditional expectation. 
\begin{definition} Under Assumptions \ref{assumption:bounded bilinear forms} and \ref{assumption:carleman}, the following member of $\ca H$, 
\begin{align}\label{eq:definition weak conditional mean}
    \mu \lo V + \lambda \lo {R\lo {UV} \hii c}(U) - E [\lambda \lo {R\lo {UV}\hii c}(U)],  
\end{align}
  is called the (centered version of) weak conditional expectation of $V$ given $U$ with respect to $\mathfrak M$, written as $E (V \bbar U)$.   
\end{definition}
While it would be more rigorous to use the notation $E  \hi{(\mathfrak M)}(V\bbar U)$, the Hilbert space $\mathfrak M$ is often obvious from context, and thus we omit the superscript. 

Weak conditional expectation is an extension of conditional expectation of Hilbert space-valued random elements: if $\mathfrak M$ is $L \lo 2 (P \lo U)$, then it reduces to the usual conditional expectation $E(V|U)$, as defined via Radon-Nykodim derivative. When $\mathfrak M$ is a proper subset of $L \lo 2 (P \lo U)$, the weak conditional expectation can be interpreted in the following equivalent way: for each $h \in \ca H$, $\langle h, E(V \bbar U) \rangle \lo {\ca H}$ is the $L \lo 2$-projection on to the linear space spanned by the set $\{1\} \cup \{ f(U): f \in \mathfrak M \}$. This equivalence is made rigorous in the next proposition. 
\begin{proposition}\label{proposition:equivalent definition} Suppose Assumptions \ref{assumption:bounded bilinear forms} and \ref{assumption:carleman} are satisfied. Then the following statements are equivalent: 
\begin{enumerate}
    \item $\Psi(U)$ is the weak conditional expectation $E(V\bbar U)$ with respect to $\mathfrak M$; \vspace{-.08in}
    \item  for all $h \in \ca H$ and all $m \in \mathfrak M$, 
    \begin{align}\label{eq:2 equations}
      \ali   E ( \langle h,  V - \Psi (U) \rangle \lo {\ca H} ) = 0,  \quad   E [( \langle h,  V - \Psi (U)  \rangle \lo {\ca H}  ) m (U) ] = 0. 
    \end{align}
\end{enumerate}
\end{proposition}
By the uniqueness of Riesz representation, the first equation in (\ref{eq:2 equations}) is equivalent to $E(V-\Psi(U))=0$. Furthermore, if $\mathfrak M$ is an RKHS generated by a kernel $\ka (\cdot, \cdot)$, then the left-hand side of the  second equation in (\ref{eq:2 equations}) can be rewritten as 
\begin{align*}
 E [( \langle h,  V - \Psi (U) \rangle \lo {\ca H} ) \langle m, \ka (\cdot, U) \rangle \lo {\mathfrak M} ]  = \langle h, E[ ( V - \Psi (U) ) \otimes \ka (\cdot, U) ] m \rangle \lo {\ca H}=0. 
\end{align*}
Since this holds for all $h \in \ca H$ and $m \in \mathfrak M$, the second equation in (\ref{eq:2 equations}) is equivalent to the operator  $E[ ( V - \Psi (U) ) \otimes \ka (\cdot, U) ]$ being 0. This is an easy criterion for checking whether a random element is a weak conditional expectation, and will be used in our construction. We record it as the next corollary. 
\begin{corollary}\label{corollary:2 equations} Suppose Assumptions \ref{assumption:bounded bilinear forms} and \ref{assumption:carleman} are satisfied,  and $\mathfrak M$ is an RKHS generated by a positive $\ka: \ca G \times \ca G \to \real$. Then $\Psi (U ) = E (V \bbar U )$ if and only if 
\begin{align*}
 E[  V - \Psi (U)  ]=0, \quad   E[ ( V - \Psi (U) ) \otimes \ka (\cdot, U) ] = 0.  
\end{align*}
\end{corollary}

\def\var{\mathrm{var}}

\subsection{Weak Bayes rule for global Fr\'echet regression}

In this subsection we show that, if we replace the posterior $E  [ E  (d \hi 2 (Y, y)  | c, f ) | U \lo {1:n}(y) ]$ by its weak counterpart $E  [ E  (d \hi 2 (Y, y)  | c, f ) \bbar  U \lo {1:n}(y) ]$, then we arrive at the same estimates given in Theorem \ref{theorem:posterior means for c and f} without using the Gaussian prior in Assumption  \ref{assumption:Bayesian Frechet Gaussian prior}.
We replace the Assumption  \ref{assumption:Bayesian Frechet Gaussian prior}  by the following moment conditions: 
\begin{assumption}\label{assumption:weak Bayes assumption} \ 
\begin{enumerate}
\item     $E[U \lo x(y) | c, f]=   c + \langle f, \ka \lo c (\cdot, x) \rangle \lo {\ca H}, \quad \var[ U \lo x (y)  | c, f] = \sigma \lo y \hi 2$; \vspace{-.04in}
\item  $\cov[U \lo i(y), U \lo j (y) | f, c] = 0$ for all $i \ne j$, and $\cov[U \lo x(y), U \lo j (y) | f, c] = 0$ for all $j$; 
 \vspace{-.04in}
 \item $E(c) = \alpha \lo y$, $\var (c) = \tau \lo y \hi 2$ for some $\alpha \lo y, \tau \lo y \in \real$;  \vspace{-.04in}
\item    $ E(f) =   \nu \lo y, \quad \var ( f) = \Lambda \lo y, \quad \cov(f,c)=0$.  
\end{enumerate}
\end{assumption}
There is no Gaussian assumption or any distributional assumption, but only assumptions about the first two moments. Moreover, independent assumptions are replaced by the weaker conditions of uncorrelatedness.  

We next use the weak conditional mean to define the weak version of the Fr\'echet Bayes rule, where the posterior means $E(c | U \lo {1:n}(y))$ and $E(f | U \lo {1:n}(y))$ are replaced by weak posterior means $E(c \bbar U \lo {1:n}(y))$ and $E(f \bbar U \lo {1:n}(y))$. We first identify the three basic components needed for defining the weak condition expectation in our context: $\ca G$, $\mathfrak M$, and $\ca H$. For both $c$ and $f$, $\ca G$ is the metric space $(\Omega \lo Y, d)$, and $\mathfrak M$ is the linear space 
\begin{align}\label{eq:frak M}
     \{ b \lo 0 + b \lo 1 U \lo 1 (y) + \cdots + b \lo n U \lo n (y): b \lo 0, \ldots, b \lo n \in \real \}. 
\end{align}
For $c$, the space $\ca H$ is simply $\real$, and for $f$, the space $\ca H$ is the RKHS   spanned by $\{\ka (\cdot, x \lo i): i = 1, \ldots, n \}$, which is also denoted by $\ca H$.  

\begin{theorem}\label{theorem:weak Bayes rule} Under Assumption \ref{assumption:weak Bayes assumption}, the weak posterior expectation of $c$ and  $f$  with respect to the space (\ref{eq:frak M})  are
\begin{align*} 
E  (c\bbar  U \lo {1:n}(y))
= \ali 
\frac{\sigma \lo y \hi 2 \alpha_y+n\tau_y^2\bar U (y) }
{\sigma \lo y \hi 2 +n\tau_y^2}, \\
E  (f \bbar U \lo {1:n}(y))
= \ali 
\left(\Lambda_y \hi \dagger +\sigma \lo y^{-2}B^\ast B\right) \hi \dagger
\left(
\Lambda_y\hi \dagger \mu+\sigma \lo y^{-2}B^\ast U \lo {1:n} (y)
\right).
\end{align*}
\end{theorem}

We next turn to the weak posterior predictive mean $E[ E ( U \lo x (y)| c, f) \bbar U \lo {1:n}(y)] $. As one might expect, the relation $E[ E ( U \lo x (y)| c, f) | U \lo {1:n}(y)] = E[  U \lo x (y) | U \lo {1:n}(y)]$ has its parallel form in the weak conditional mean setting, as stated in the next proposition. 
\begin{proposition} Under Assumption \ref{assumption:weak Bayes assumption}, and  with respect to $\mathfrak M$ defined in (\ref{eq:frak M}), 
\begin{align*}
    E[ E ( U \lo x (y) | c, f) \bbar U \lo {1:n}(y)] =     E[  U \lo x (y)   \bbar U \lo {1:n}(y)]. 
\end{align*}
\end{proposition}

We now give the formal definition of the weak Fr\'echet Bayes rule. 
\begin{definition}
    Suppose $Y$  is a member of $F \lo d \hi 2 (P)$.   Then the weak Fr\'echet Bayes (WFB) rule for $E \lo \oplus (Y \lo x |\theta) $ based on the sample $Y \lo 1, \ldots, Y \lo n$, is  
\begin{align}\label{eq:objective function for weak Bayes rule}
\hat Y \lo {\mathrm{WFB}} = \underset{y \in {\Omega \lo Y }}{\argmin}  \,   E [U \lo x (y)    \bbar  U \lo {1:n} (y)  ].  
\end{align}    
\end{definition}

\medskip 

We call the quantity $E[  U \lo x (y)   \bbar U \lo {1:n}(y)]$ the weak posterior predictive mean of $U \lo x (y)$. 
  The next corollary gives the explicit expression of $E [  U \lo x (y)  \bbar U \lo {1:n}(y)] $. 

\begin{corollary} Under Assumption \ref{assumption:weak Bayes assumption}, we have 
\begin{align*} 
E [  U \lo x (y)   \bbar U \lo {1:n}(y)] = E(c \bbar U \lo {1:n}(y) ) + \langle \ka \lo c (\cdot, x), E ( f \bbar U \lo {1:n}(y)) \rangle \lo {\ca H}, 
\end{align*}
where $E(c \bbar U \lo {1:n}(y) )$ and 
$E  (f \bbar U \lo {1:n}(y))$ are as given in Theorem \ref{theorem:weak Bayes rule}. 
\label{corollary:waek posterior predictive mean}
\end{corollary}

\def\eop{\hfill $\Box$}

Our approach of weak Bayes rule aligns with the principles of Bayesian linear inference \citep{hartigan_linear_1969,goldstein_linear_1980, mouchart_note_1984}. The involvement of the Hilbert space-valued $f$ requires the use of the weak conditional mean. This shift not only justifies the use of the previously derived estimators under non-Gaussian noise but also connects our framework to the broader literature on robust Bayesian modeling \citep{berger_overview_1994}, where the posterior is viewed as a coherent update of beliefs based on second-order structure rather than full likelihood specifications.

\section{Implementation}\label{sec:implement}

Section \ref{subsec:bfr.prior} details the prior specification and coordinate representation, and Section \ref{subsec:bfr.bfr} provides implementation strategies and an algorithm.

\subsection{Prior specification and coordinate representation}\label{subsec:bfr.prior}

The centered  Bayesian nonlinear regression model in Assumption \ref{assumption:Bayesian Frechet Gaussian prior} (or Assumption~\ref{assumption:weak Bayes assumption}) is parameterized by five prior parameters: the noise variance $\sigma \lo y \hi 2 $, the prior mean $\alpha_y\in\bbR$ and variance $\tau^2 \lo y$ of the intercept $c$,  the mean function $\nu_y\in\ca H$ and covariance operator $\Lambda \lo y:\ca H\to\ca H$ of the regression functional component $f$. 

We specify the covariance operator $\Lambda \lo y$ as a rescaled version of the empirical covariance operator of $X$: $\Lambda \lo y=\omega \lo y^2 \Sigma_{XX}^\gamma$ for some power $\gamma\in\bbR$.
Let $\{(\lambda_i, \phi_i)\}_{i=1}^{n}\subset \bbR\times \ca H$ denote the eigen-pairs of the empirical covariance operator $\Sigma_{XX}=\sum_{i=1}^{n}\lambda_i(\phi_i\otimes \phi_i)$. Then, we have $\Sigma_{XX}^\gamma = \sum_{i=1}^{n} \lambda_i^\gamma (\phi_i\otimes \phi_i)$, where we adopt the convention that $\lambda_i^\gamma=0$ whenever $\lambda_i=0$ regardless of the sign of $\gamma$. This specification encompasses several seminal prior structures. The case  $\gamma=-1$ yields a functional analogue of Zellner's $g$-prior \citep{zellner_assessing_1986}, i.e., $\Lambda \lo y = \omega \lo y^2\Sigma_{XX}^\dagger$, which is related to the statistical information. The case $\gamma=0$ recovers the identity operator, which is closely related to a Gaussian process regression with   $f\sim\GP(\nu \lo y, \sigma^2 \lo \omega \kappa )$. For $\gamma >0$, the prior follows the structure proposed by \cite{krishna_bayesian_2009} to address high-dimensional collinearity, a common challenge in the analysis of complex objects. Under this operator-based specification, a Gaussian prior on $f$ admits a Karhunen-Loève-style basis expansion: $f = \nu_y + \sum_{i=1}^{n} z_i\lambda_i^{\gamma/2} \phi_i$, where $ z_i\overset{\iid}{\sim} N(0, \omega \lo y^2)$.

Note that the posterior predictive mean in \cref{eq:bfr.nlbr.ppm_rkhs} depends on the prior variances only through the ratios $r^2_\tau=\tau \lo y^2/\sigma \lo y \hi 2 $ and $r_\omega^2=\omega \lo y^2/\sigma \lo y \hi 2 $. Thus, we omit the specification of $\sigma \lo y \hi 2 $, $\tau \lo y \hi 2$, and $\omega \lo y \hi 2$ in favor of these relative variance ratios $r_\tau^2$, $r_\omega^2>0$. 
While the noise level $\sigma \lo y \hi 2 $, $\tau \lo y \hi 2$, and $\omega \lo y \hi 2$ likely vary with the choice of $y$, the variance ratios can be held constant across the metric space, as they represent the strength of the prior belief relative to the data-driven evidence.

The prior mean parameters are specified as follows. Given a guess or prior estimate of the regression function evaluated at $x_1,\ldots, x_n$
\begin{align*}
 \eta \lo y = (    E \{d \hi 2 (Y \lo 1, y)\}, \ldots,  E \{d \hi 2 (Y \lo n, y) \} ) \trans, 
\end{align*}
we then define the prior intercept as the empirical average $\bar \eta \lo y=\sum \lo {i=1}\hi n \eta \lo {yi}   / n$. Then, the hyperparameter $(\nu \lo y(x \lo 1), \ldots, \nu \lo y (x \lo n))\trans$ is defined as the centered vector $\eta \lo y - \bar \eta \lo y \bone \lo n$. 
As we will see in Theorem~\ref{thm:post.pred.mean}, we only need to specify the evaluation of the prior mean function $\nu_y$ at $x_1,\ldots, x_n$, rather than the whole function.
Since the vector $\eta \lo y$ depends on the response object $y$, it must be updated during the minimization of the objective in \cref{eq:objective function for weak Bayes rule} to compute the posterior predictive Fréchet mean. This allows for flexible, problem-specific specification. For example, $\eta \lo y$ could be derived from a Fréchet regression on an auxiliary dataset or, in longitudinal studies, from previously observed temporal states.

Given this choice of prior, determined by $r_\tau^2$, $r_\omega^2$, $\eta \lo y$, and the power parameter $\gamma$, we next derive an alternative form of  Theorem \ref{theorem:posterior means for c and f} and Corollary \ref{corollary:E U x y} (equivalently, Theorem~\ref{theorem:weak Bayes rule} and Corollary~\ref{corollary:waek posterior predictive mean}) in terms of  Euclidean vectors and matrices rather than Hilbertian functions and linear operators via coordinate representation (see, for example, \cite{li_sufficient_2018} and \cite{li_nonparametric_2018}).  Let  $K=[\kappa (x_i,x_j)]_{i,j=1}^{n}$ be the $n\times n$ Gram matrix. Then  $B \hi * B =H K H \equiv G$, where $ H= I \lo n - \bone \lo n \bone \lo n^\top/ n $.   Let $\{(\lambda_i, v_i)\}_{i=1}^n$ denote the eigen-decomposition of $n \inv G$, with eigenvalues in decreasing order (i.e., $\lambda_1 \ge \lambda_2 \ge \dots \ge \lambda_n$). Let $s$ denote the number of nonzero eigenvalues, and then $B \hi * B =n\sum_{i=1}^{s}\lambda_i v_i v_i^\top$. For a target $x\in{\Omega \lo X} $, we define the centered $n$-dimensional kernel vector $  k \lo c (x)$ be the vector
 $ 
   H[ (\ka   (x, x \lo 1), \ldots \ka (x, x \lo n) ) \trans -  K\bone \lo n/n]. 
$

\begin{theorem}\label{thm:post.pred.mean}
Under Assumption \ref{assumption:Bayesian Frechet Gaussian prior} and the prior specifications that
\begin{align*}
\Lambda \lo y =\omega \lo y ^2\Sigma_{XX}^\gamma, \ 
r_\tau^2=\tau \lo y \hi 2/\sigma \lo y \hi 2, \ 
r_\omega^2=\omega \lo y^2/\sigma \lo y \hi 2, \ 
\eta \lo y - \bone \lo n \bar \eta \lo y= ( \nu \lo y (x \lo 1), \ldots, \nu \lo y (x \lo n ) ) \trans, \ \textit{and} \
\alpha \lo y = \bar \eta \lo y,
\end{align*} 
the posterior predictive mean of $U \lo x (y)$, expressed in Euclidean vectors and matrices,  is 
\begin{align}
\begin{split}
\ali  E [ U \lo x (y) | U \lo {1:n}(y)] \\
\ali =
    \frac{\bar \eta \lo y 
    /r_\tau^2 +n  \bar U (y)   }{1/r_\tau^2 + n}
    + k \lo c (x)^\top
    \left(
        \sum_{i=1}^{s}\frac{r^2_\omega\lambda_i^\gamma}{nr_\omega^2\lambda_i^{1+\gamma}+1} v_i v_i^\top
    \right)
    \left(
        U \lo {1:n} (y)  + \sum_{i=1}^{s}
        \frac{v_iv_i^\top \eta \lo y}{n r_\omega^2\lambda_i^{1+\gamma}}
    \right).\label{eq:bfr.prior.ppm1}
\end{split}
\end{align}
\end{theorem}

 \medskip 

This formulation offers the flexibility of truncating the eigen-decomposition at some $s \ll n$, which is particularly useful for handling large datasets. Such truncation not only reduces computational overhead but also improves numerical stability by effectively regularizing the spectral representation and discarding noise-dominated eigen-components. In practice, $s$ may be selected to capture a specified large proportion of the total empirical variance. 
We refer to Section~\ref*{sec:supp_synthetic} of the Supplementary Materials for numerical experiments investigating the effect of truncation on prediction performance.

Under a functional Zellner’s $g$-prior ($\gamma = -1$), as shown in the following corollary, the predictive mean simplifies significantly:
\begin{corollary}
    If $\gamma=-1$, the posterior predictive mean function is given by:
    \begin{align}
     E [U \lo x (y)|U \lo {1:n}(y)]
        &=
        \frac{\bar \eta \lo y/r_\tau^2 + n\bar U  (y) }{1/r_\tau^2 + n}
        + k \lo c (x)^\top
        G^\dagger
        \frac{\eta \lo y/r_\omega^2 + n U \lo {1:n}(y) }{1/r_\omega^2+n}, \label{eq:bfr.prior.ppm_zellner}
    \end{align}
    where $\dagger$ denotes the Moore-Penrose inverse of a matrix. 
\end{corollary}

Note that if we choose $r^2_\omega=r_\tau^2=r^2$, the posterior predictive mean becomes equivalent to a frequentist nonlinear global Fréchet regression, where the response vector is replaced by a convex combination of the data and the prior: $U \lo {1:n}(y) \leftarrow w U \lo {1:n}(y)  + (1-w) \eta \lo y$, with the weight $w=(1+1/nr^2)^{-1}$ depending on the variance ratio. As $n\to\infty$ or $r^2\to\infty$, we have $w\to 1$, recovering the frequentist estimate, and conversely, as $r^2\to 0$, the weight $w\to 0$ such that the estimator is driven entirely by the prior. This behavior provides a functional analogue to the geodesic Stein shrinkage for Fréchet means \citep{mccormack_stein_2022}, extending the concept of shrinkage from a single point to an entire regression surface.

\subsection{Algorithm}\label{subsec:bfr.bfr}

The practical realization of Bayesian Fréchet regression entails minimizing the posterior predictive mean \cref{eq:bfr.prior.ppm1} over the target response $y\in\Omega \lo Y$ to obtain the posterior predictive Fréchet mean \cref{eq:objective function for weak Bayes rule} at some $x\in{\Omega \lo X} $. This optimization necessitates repeated evaluations of $E [U \lo x (y) |U \lo {1:n}(y)]$ at various $y$ and $x$ values. The overall procedure is summarized in Algorithm~\ref{alg:bnlrmn}. 
\begin{algorithm}[ht]
\caption{The Proposed Bayesian Nonlinear Regression of the Squared Metric
\label{alg:bnlrmn}}
\KwIn{Metric space $(\Omega \lo Y, d )$, training data $ \{(x_i, Y_i)\}_{i=1}^n\subset {\Omega \lo X} \times \Omega \lo Y$, kernel $\kappa$ over ${\Omega \lo X} $, variance ratios $r_\tau^2, r_\omega^2>0$, prior guesses $\eta \lo y\in\bbR^n$,  
covariance prior power $\gamma$, target features $x\in{\Omega \lo X} $ and target response $y\in\Omega \lo Y$.}
\KwOut{Posterior predictive mean $E[U \lo x(y) | U \lo {1:n}(y)]$.}
\BlankLine
\tcp{Phase I: Pre-Computation (Execute Once)}
Construct the centered Gram matrix $G=HKH$, where $K=[\kappa(x_i, x_j)]_{i,j}$\;
Compute the eigendecomposition of $n\inv G$ to obtain $\{(\lambda_i,  v_i)\}_{i=1}^{n}$, in descending order of eigenvalues. Let $s$ denote the number of nonzero eigenvalues.\;
Pre-calculate the eigenvalues of $r^2_\omega\lambda_i^\gamma/(n r_\omega^2\lambda_i^{1+\gamma}+1)$ and $1/n r_\omega^2\lambda_i^{1+\gamma}$ to obtian the spectral weights in \cref{eq:bfr.prior.ppm1}.

\BlankLine
\tcp{Phase II: Evaluation (Execute for each $x, y$)}
\ForEach{$x,y$}{
Compute the centered feature vector $[k(x) - n\inv  K\bone \lo n]^\top H$, where $k(x) = [\kappa(x, x_i)]_{i=1}^{n}$\;
Compute the vector of responses $U \lo {1:n}(y) =[U_i(y)]_{i=1}^{n}$, where $U_i(y)=d ^2(Y_i, y)$\;
Compute $E [U \lo x(y) |U \lo {1:n}(y)]$ using \cref{eq:bfr.prior.ppm1}.
}
\Return
\end{algorithm}
From an algorithmic perspective, a key feature of our framework is that the most intensive operation to compute the eigen-decomposition of the Gram matrix is performed only once as a pre-processing step. For a fixed $x$, the evaluation of the objective function for different $y$ requires only the computation of the distances $\{d ^2(Y \lo i, y)\}_{i=1}^n$ and $\{d ^2(\eta \lo {yi}, y)\}_{i=1}^n$, followed by efficient matrix-vector products. The hyperparameters $r_\tau^2$ and $r_\omega^2$ influence the variational objective solely through the scalar spectral weights, leaving the eigenvectors invariant. This invariance permits immediate sensitivity analyses and cross-validation over a wide grid of prior-to-noise variance ratios by simply re-scaling the pre-computed eigenvalues. Furthermore, this structure facilitates a highly efficient solution path approach: when optimizing over a sequence of monotonic variance ratios, the Fréchet minimizer from the previous parameter state can be used to warm-start the optimization for the next, vastly accelerating convergence across the entire hyperparameter space. 

The final step in operationalizing this framework is the minimization over $y\in\Omega \lo Y$, which is inherent to Fréchet regression. While the specific optimization strategy depends on the geometry of the metric space $(\Omega \lo Y, d )$, our Bayesian objective \eqref{eq:bfr.prior.ppm1} possesses several desirable properties. Because $E[U \lo x (y) |U \lo {1:n}(y)]$ is an affine combination of squared distances to the training responses and prior guesses, it inherits the regularity of the squared metric. In Euclidean spaces or on Riemannian manifolds, where $d \hi 2 $ is typically differentiable, gradient-based or Riemannian steepest descent methods are directly applicable, provided the prior mean is smoothly specified. In more complex settings, such as Wasserstein spaces where $y$ represents a probability distribution, the minimization can be performed via the discretization of the quantile function, transforming the task into a tractable constrained optimization problem \citep{petersen_frechet_2019, bhattacharjee_nonlinear_2025}.

\section{Simulation Studies}\label{sec:sims}

We consider two simulation studies to illustrate the empirical performance, shrinkage behavior, and robustness of the proposed Bayesian Fréchet regression (BFR) framework. Operationally, implementing BFR requires specifying five core components: a metric space $(\Omega \lo Y, d)$, a training dataset $\cD = \{(Y_i, x_i)\}_{i=1}^{n}\subset \Omega \lo Y\times {\Omega \lo X} $, a positive-definite kernel $\ka  $ over the predictor space ${\Omega \lo X} $, and a prior mean vector $\eta \lo y$ with entries representing the guesses of $E \{d ^2(Y, y)\mid X=x_i\}$, and the prior-to-noise ratios $r_\tau^2$ and $r_\omega^2$. Given these ingredients, the Fr\'echet Bayes rule at a target location $x\in{\Omega \lo X} $ is obtained by executing the metric-specific minimization strategy detailed in Sections~\ref{sec:bfr} and \ref{sec:implement}.

\subsection{Distributional responses in Wasserstein space}\label{subsec:sims.gaussian}

In our first numerical experiment, we let $\Omega \lo Y$ be the space of univariate Gaussian distributions on $\bbR$, parameterized by the mean $\xi\in\bbR$ and standard deviation $\delta$. This space is equipped with the 2-Wasserstein distance $d_{W_2}$, which admits a closed-form expression: for two Gaussian densities $Y = N(\xi, \delta  \hi 2 ),   Y' = N(\xi', \delta'^2) \in \Omega \lo Y$, the squared 2-Wasserstein distance is given by $d_{W_2}^2(Y, Y') = (\xi-\xi')^2 + (\delta - \delta')^2$. Geometrically, this structure renders $(\Omega \lo Y, d_{W_2})$ isometric to the Euclidean upper half-plane $\bbR\times \bbR_{>0}$. We generate a single predictor $X$ uniformly distributed over $[-2, 2]$ and establish a linear relationship for the distribution parameters:
$\xi = X + 0.3\eps_\xi$ and $\delta = 3+0.3x+0.2\eps_\delta$, where $\eps_\xi,\eps_\delta\sim N(0,1)$. Under this example, the true conditional expected squared distance can be analytically derived. For an arbitrary target distribution $Y=N(\xi,\delta)$, the conditional expectation is given by:
\begin{align*}
    E \{d \hi 2 (Y, y) \mid X=x\} = 0.2^2+0.3^2+(\xi-x)^2 + (\delta-3-0.3x)^2,
\end{align*}
where $0.13 = 0.3^2 + 0.2^2$ represents the irreducible noise variance in the metric space.

We utilize this analytical form to construct the vector of prior guesses $\eta \lo y$, setting $\eta \lo {yi} = E [d \hi 2 (Y, y)  \mid X=x_i]$.  To examine the shrinking behavior of the estimator, we vary the variance ratios across an isotropic grid ($r_\tau \hi 2=r_\omega \hi 2$) and fix the prior covariance power to the functional Zellner specification ($\gamma=-1$). The predictor space is equipped with a Gaussian radial basis kernel over ${\Omega \lo X} =\bbR$, that is, $\ka  (x, x') \propto \exp\{-(x-x')^2/2h^2\} $, where the bandwidth $h$ is chosen via the standard median heuristic over the training sample (see \cite{scholkopf2002learning}).

\begin{figure}{H}
    \centering
    \includegraphics[width=\linewidth]{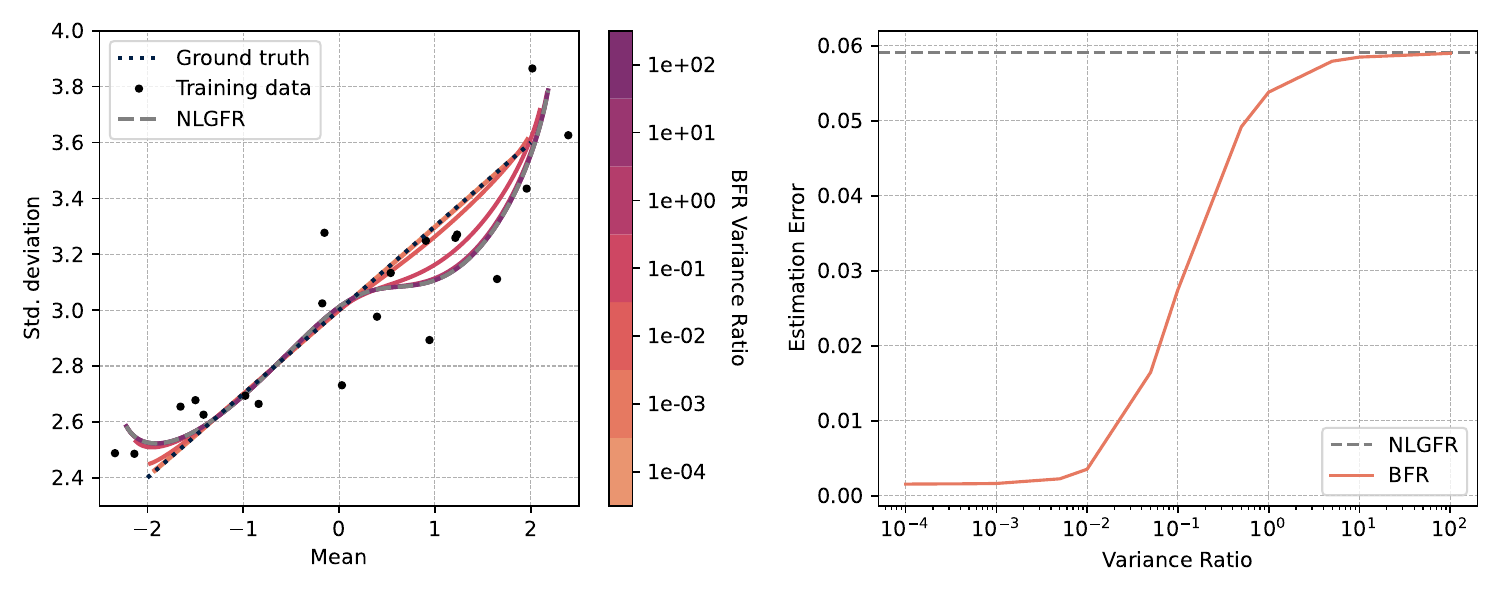}
    \caption{Bayesian global Fréchet regression for univariate Gaussian distributional responses. Left Panel: The true conditional Fréchet mean function compared against the estimated posterior predictive Fréchet mean functions under the proposed BFR framework across a path varying variance ratios, alongside the frequentist NLGFR estimator.
    Right Panel: The integrated squared Wasserstein distance between the estimated posterior predictive Fréchet mean and the true conditional Fréchet mean as a function of the prior-to-noise variance ratio, illustrating the controlled interpolation between the NLGFR estimator and the prior.}
    \label{fig:sims.gaussian}
\end{figure}

We simulate $n=20$ training points to highlight the regularizing capability of the prior. The posterior predictive Fréchet mean is evaluated over a fine grid over the domain $[-2,2]$. To solve the variational minimization over $y\in\bbR^2$, we utilize gradient-based optimization. Figure~\ref{fig:sims.gaussian} illustrates the trajectory of the estimated posterior predictive Fréchet mean function across a path of variance ratios, along the corresponding average distance to the true conditional Fréchet mean function. We also include the frequentist NLGFR estimator of \cite{bhattacharjee_nonlinear_2025} as a baseline, which corresponds asymptotically to the limit $r_\tau \hi 2=r_\omega \hi 2=\infty$. Varying the variance ratios from $\infty$ toward $0$ increases the strength of the prior, orchestrating a smooth geometric interpolation between the NLGFR estimator and the prior. Notably, as the variance ratio approaches $0$, the posterior predictive Fréchet mean recovers the true conditional Fréchet mean, subject only to a minimal approximation error induced by the finite-dimensional truncation of the kernel embedding.

\subsection{Spherical responses on the unit sphere}\label{subsec:sims.sphere}

In our second numerical experiment, we evaluate the proposed framework on a non-linear Riemannian manifold with positive curvature, adopting the spherical response setting introduced by \cite{petersen_frechet_2019}. Specifically, we consider response objects constrained to the unit sphere $\bbS^2\subset\bbR^3$, equipped with the geodesic distance $d(y, y')=\arccos(y^\top y')$. A single predictor $X$ is generated uniformly over the unit interval $[0, 1]$, and the true   Fréchet mean function is defined as the spherical trajectory:
\begin{align*}
    E \lo \oplus (Y \lo x) = 
    \left(
    (1-x^2)^{1/2}\cos(\pi x),
    (1-x^2)^{1/2}\sin(\pi x),
    x
    \right)^\top
\end{align*}
To simulate realistic non-Euclidean noise, random response objects are generated by adding isotropic Gaussian noise in the ambient space $\bbR^3$ centered at $E \lo \oplus (Y \lo x)$ and subsequently projecting the perturbed vectors back onto the sphere $\bbS^2$.

The predictor space is equipped with the squared-exponential kernel over ${\Omega \lo X} =\bbR$, whose bandwidth is calibrated via the empirical median heuristic. To construct a vector prior guesses $\eta \lo y$, we sample $50$ auxiliary pseudo-responses distributed around the true spherical trajectory and compute their empirical average squared geodesic distance to $y\in\bbS^2$. To optimize over $y\in\bbS^2$, we implement a projected gradient descent algorithm on the manifold.

\begin{figure}[H]
    \centering
    \includegraphics[width=\linewidth]{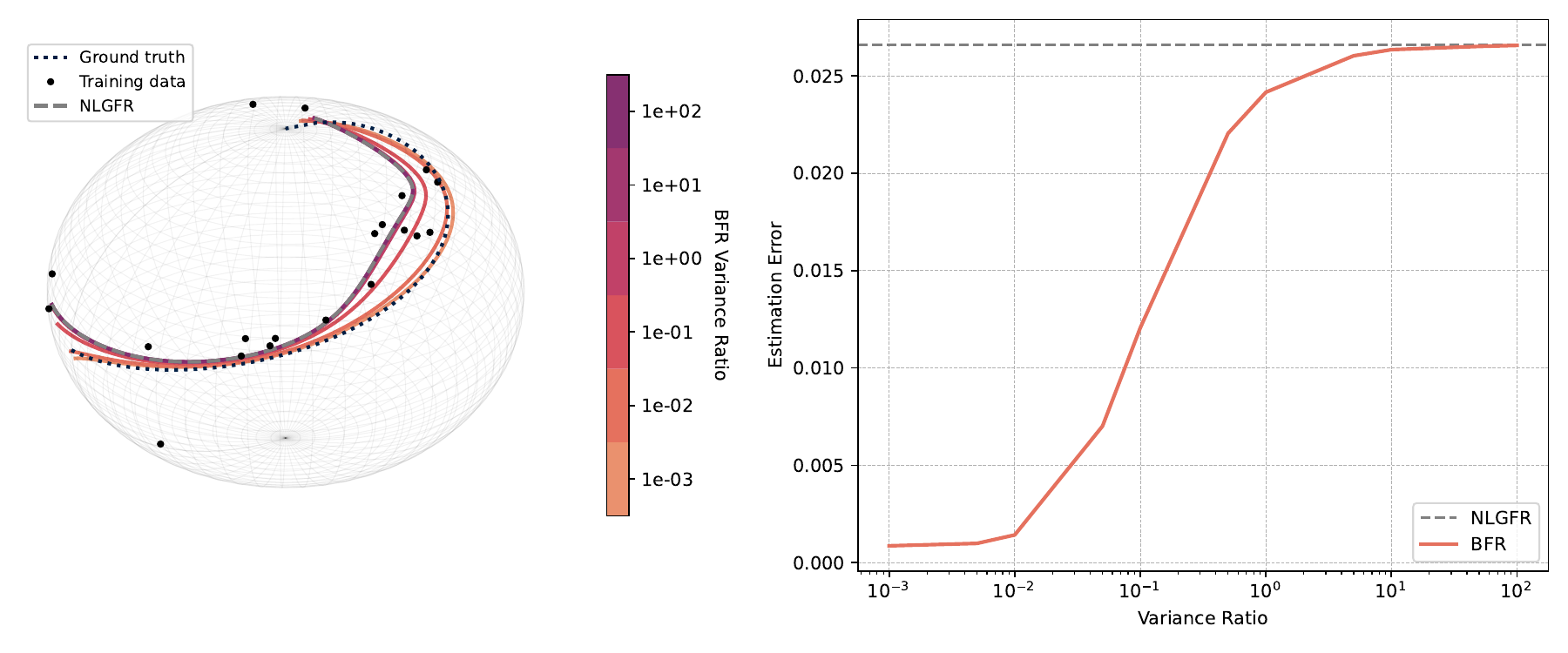}
    \caption{Bayesian global Fréchet regression for spherical responses on the unit sphere. Left Panel: The true conditional spherical trajectory compared against the estimated posterior predictive Fréchet mean functions generated by BFR across a regularization path of variance ratios, alongside the frequentist NLGFR estimator. Right Panel: The integrated squared geodesic distance between the estimated Fréchet mean functions and the true conditional surface as a function of the prior-to-noise variance ratio.}
    \label{fig:sims.sphere}
\end{figure}

Figure~\ref{fig:sims.sphere} displays the estimated posterior predictive Fréchet mean trajectories across the sphere alongside their corresponding estimation errors. The empirical results mirror the regularization dynamics observed in the Wasserstein setting: the proposed BFR framework acts as a structurally sound geometric bridge, interpolating smoothly between the frequentist NLGFR baseline  ($r^2 = \infty$) and the prior ($r^2 \to 0$). Despite the positive curvature and non-linear constraints, the variational inference step successfully maps the linear updates from the RKHS predictor space back onto the manifold surface without boundary violations. As the variance ratio decreases, the estimator steadily approaches the true conditional trajectory, bounded only by the inherent approximation error of the finite-rank kernel representation. This confirms that the computational and regularizing advantages of our framework hold across fundamentally distinct, non-Euclidean geometric structures.

To examine the robustness of our framework under model misspecification, we conduct an additional simulation study where the prior structurally departs from the true data-generating process. Specifically, we rotate the true mean function around the $z$-axis by a varying angle. As the angle increases, the prior becomes progressively misspecified, so the BFR estimator will be shrinking towards an incorrect mean function. Within this adversarial setup, we evaluate three configurations of the prior covariance power: $\gamma\in\{-1,0,1\}$, to study how the spectral geometry of the prior modulates robustness to misspecification.

\begin{figure}[H]
    \centering
    \includegraphics[width=\linewidth]{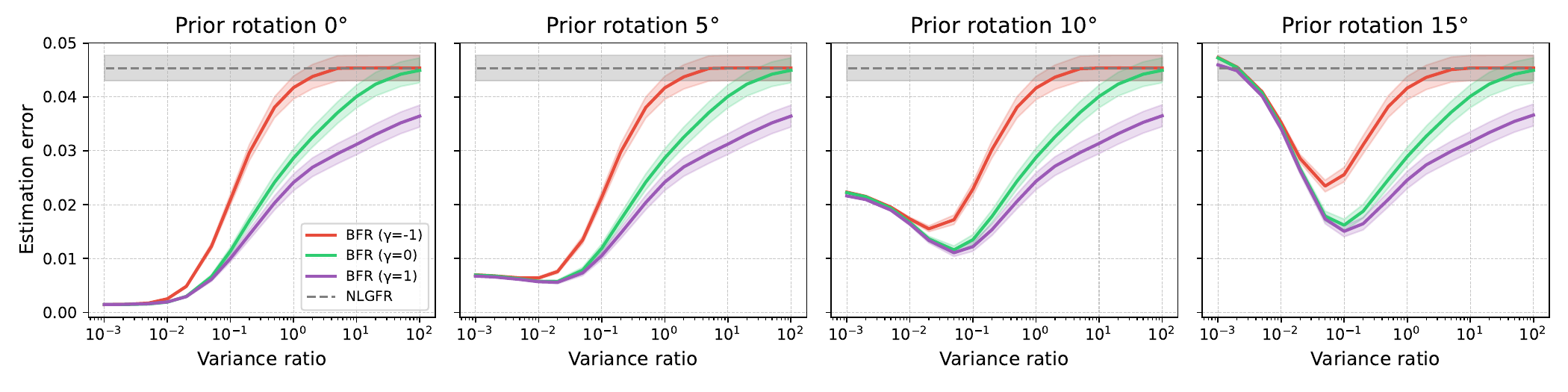}
    \caption{Prior misspecification and robustness analysis for spherical responses on the unit sphere. The integrated squared geodesic distance between the estimated posterior predictive Fréchet mean and the true conditional Fréchet mean, evaluated across $100$ independent Monte Carlo replications. Panels compare the estimation risk across a gradient of prior misalignment induced by an angular rotation around the $z$-axis. Curves display the empirical mean $\pm$ one standard error for the proposed BFR framework under three spectral prior covariance configurations against the NLGFR baseline.}
    \label{fig:sims.sphere_rotated}
\end{figure}
 
Figure~\ref{fig:sims.sphere_rotated}  details the integrated estimation error as a function across the misspecification angles and the variance ratios. Under mild misspecification (i.e., small rotations), highly regularized models (i.e., a small $r^2$) outperform the NLGFR baseline, as the regularization benefit outweighs the minor bias introduced by the prior. However, under severe misspecification (i.e., large rotations), the structural bias dominates, rendering an intermediate variance ratio optimal. This phenomenon is analogous to classical Stein shrinkage estimation, where the shrinkage factor (similar to our variance ratios) determines the risk reduction. 

Moreover, as for the choice of the prior covariance power $\gamma$, while the identity prior ($\gamma = 0$) and the hyper-regularized prior ($\gamma = 1$) achieve comparable minimum estimation errors at their optimal hyperparameter tunings, Zellner's $g$-prior ($\gamma = -1$) exhibits pronounced vulnerability to severe misspecification. Because Zellner’s specification forces the prior covariance to mimic the inverse structural variance of the design matrix, it restricts deviations from the prior precisely along the principal directions of covariate variation. When the prior is misspecified, the $g$-prior lacks the directional flexibility necessary to adaptively recover the true underlying surface from data, highlighting a key trade-off between classical Bayesian efficiency and robust non-parametric regularization.

\section{Application to Microbiome Compositional Analysis}\label{sec:microbiome}

Recent advances in high-throughput marker-gene and metagenomic sequencing technologies have catalyzed large-scale microbiome and metagenomic studies to elucidate the structural associations between the human microbiota and complex pathophysiological states. However, a pervasive bottleneck in contemporary metagenomics is the lack of generalizability of the results across different patient populations and the poor portability of predictive models across independent studies \citep{han_techniques_2025}. We leverage the proposed Bayesian global Fréchet regression framework to address this challenge. 

To evaluate the cross-study portability, we consider the \texttt{curatedMetagenomicData} repository \citep{pasolli_accessible_2017} that consists of a standardized, large collection of human metagenomic profiles along with curated demographic and clinical data. From this database, we extracted three independent, comparable cohorts investigating the structural links between the gut microbiota and type 2 diabetes (T2D). While each study catalogs a core set of common covariates, including age, sex, body mass index (BMI), antibiotic usage, and T2D diagnosis status, the demographic profiles exhibit structural disparities:
\begin{itemize}
    \item The MetaCardis cohort \citep{forslund_combinatorial_2021} contains 234 healthy individuals and 792 individuals diagnosed with  T2D, all from European countries (mainly Germany, Denmark, and France). The cohort has a median age of 60 years, a median BMI of 29.6, balanced sex representation, and about $50\%$ antibiotic exposure.
    \item The small-scale Swedish study \citep{karlsson_gut_2013} contains 43 healthy individuals and 53 individuals with T2D, most from Sweden. This cohort consists exclusively of female participants, has a significantly higher median age of 70 years, a median BMI of 26.6, and completely excludes individuals with a recent antibiotic history.
    \item The Chinese cohort \citep{qin_metagenome-wide_2012} contains 193 healthy individuals and 170 individuals with T2D, all from China. This group has a median age of 48 years, a median BMI of 23.8, a balanced sex ratio, and no antibiotic history.
\end{itemize}
Given the well-established association of the microbiota with host genetics, epigenetics, diet, and environmental exposures (all weakly captured by the country of origin), as well as clinical factors such as age and antibiotic use \citep{cheng_gutmdisorder_2020,tang_gimica_2021,ghosh_gut_2022}, it is unclear whether the relationship between T2D and microbial composition is invariant across distinct cohorts or if it shifts due to population heterogeneity.

We consider a cross-study validation design wherein the large MetaCardis cohort \citep{forslund_combinatorial_2021} serves as an independent external training source to construct priors for the Fréchet regression between demographic and clinical variables and the microbial compositions within the two smaller target studies \citep{karlsson_gut_2013,qin_metagenome-wide_2012}. Here, the response $y\in\Omega \lo Y$ is a 101-dimensional compositional vector in the 100-dimensional open Aitchison simplex, representing the relative abundance of $K=101$ taxonomic genera under the unit-sum constraint. The predictors include age, sex, T2D status, BMI, and a binary indicator for recent antibiotic usage. To construct the prior estimate of $E \{d \hi 2 (Y, y) \mid X=x_i\}$ at a target predictor $x_i$ within a test cohort, we evaluate the fitted values from a nonlinear regression in the external MetaCardis study. We use an additive kernel structure that assigns equal weight across predictors, and for the continuous predictors (age and BMI), we use a squared-exponential kernel with bandwidths of 10 and 5, respectively. The response space is equipped with the Hellinger metric $d_{\text{Hellinger}}^2(y, y') = \frac 12 \sum_{k=1}^{K} \left(\sqrt{y_k} - \sqrt{y_k'}\right)^2$, which is well-suited for high-dimensional metagenomic abundances. 

To quantify the out-of-sample portability of the framework, given one target test cohort (\citealp{karlsson_gut_2013} or \citealp{qin_metagenome-wide_2012}), the target dataset is repeatedly partitioned into a training set, used to train the Bayesian Fréchet regression, and an evaluation set, used to evaluate the average predictive Hellinger error. By varying the slope prior-to-noise variance ratio parameter $r_\omega \hi 2$ that controls the strength of the prior, we systematically investigate whether incorporating knowledge from the external European cohort improves localized estimation compared to the frequentist NLGFR baseline \citep{bhattacharjee_nonlinear_2025}. The variance ratio for the intercept, $r_\tau \hi 2$, is set to infinity, ensuring that the external prior only regularizes the slopes without imposing a biased global mean shift.
We refer to Section~\ref{sec:supp_microbiome} of the Supplementary Materials for complete details on the experiment.

\begin{figure}[H]
    \centering
    \includegraphics[width=\linewidth]{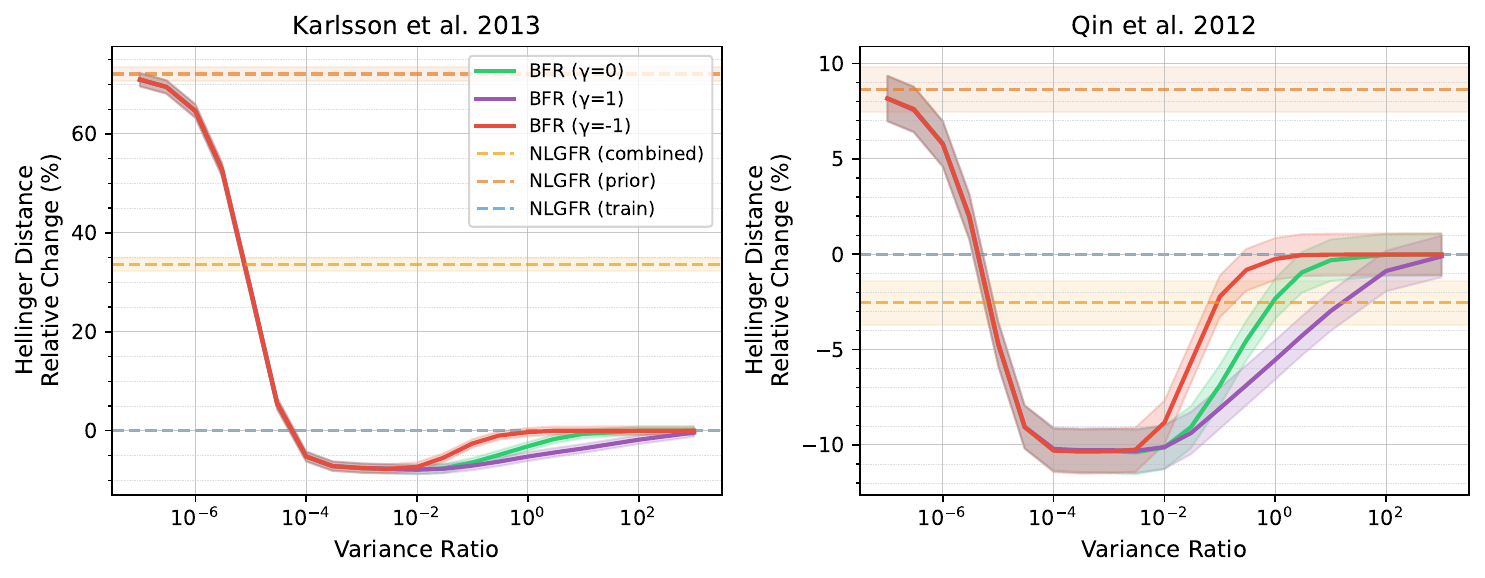}
    \caption{Cross-study predictive portability for microbiome compositional data across the Aitchison simplex. The plots display the mean prediction error, measured by the Hellinger metric and evaluated over $30$ independent replications (with shaded bands representing standard errors). The  proposed BFR model with the varying slope variance ratio parameter under three choices of the prior covariance power $\gamma$ is trained and evaluated on the target cohort
    \cite{karlsson_gut_2013} (Left Panel) or
    \cite{qin_metagenome-wide_2012} (Right Panel) using an empirical prior constructed from the external MetaCardis cohort \citep{forslund_combinatorial_2021}. The frequentist NLGFR baseline is displayed as a horizontal line. }
    \label{fig:microbiome.metrics_hellinger}
\end{figure}

Figure~\ref{fig:microbiome.metrics_hellinger} shows the predictive performance, quantified via the average Hellinger distance between the estimated posterior predictive Fréchet mean surface and the observed test profiles. We evaluate the proposed BFR framework across three distinct prior covariance powers ($\gamma=-1,0,1$) and the frequentist NLGFR baseline. As the slope variance ratio $r_\omega \hi 2$ increases, the regularizing influence of the prior decays, and the BFR profiles approach the NLGFR, verifying our theoretical results empirically. Conversely, as the variance ratio $r_\omega \hi 2$ decreases,  the prior information from the large external MetaCardis cohort is adaptively incorporated into the target estimation. Under sparse or noisy sample regimes within the target cohorts, this Bayesian shrinkage yields substantial and statistically significant drops in prediction error. This empirical result underscores the capacity of the proposed framework for cross-study portability in complex, non-Euclidean data structures.
Section~\ref*{sec:supp_microbiome} of the Supplementary Materials contains additional evaluation measures, which differ from the metric used in BFR; we find similar results to those in Figure~\ref{fig:microbiome.metrics_hellinger}.

\section{Conclusion}\label{sec:discussion}

We have introduced Bayesian global Fréchet regression as a general, operator-theoretic framework for incorporating prior structural knowledge into regression models with non-Euclidean responses. The proposed framework anchors the prior directly onto the conditional expected squared distance $E [d \hi 2 (Y, y)\mid X=x_i]$ to some arbitrary element $y$ of the metric space $\Omega \lo Y$ for all training predictors $x_i$, $i=1,\ldots, n$. The strength of this prior is adaptively controlled by variance ratio parameters. As the variance ratio varies from from to infinity, the posterior predictive Fréchet mean function orchestrates a smooth geometric interpolation, transitioning seamlessly from the regularizing prior profile to the purely data-driven frequentist nonlinear global Fréchet regression surface. Our empirical and numerical investigations validate that this Bayesian formulation yields statistically significant reductions in predictive performance compared to frequentist alternatives. This phenomenon is closely reminiscent of the Stein effect, where an appropriately chosen shrinkage factor induces risk reduction. Finally, the novel concept of Fréchet Bayes rule is introduced outside the context of Fréchet regression. So, it has the potential to become a general mechanism for Bayesian inference in metric spaces. We'll explore these potentials in future research.

\newpage

\bibliography{ref}

\dummylabel{subsec:supp_derivations.nlbr}{S1.1}
\dummylabel{subsec:supp_derivations.prior}{S1.4}
\dummylabel{subsec:supp_derivations.cr_nlbr}{S1.3}
\dummylabel{subsec:supp_derivations.gp}{S1.5}
\dummylabel{sec:supp_microbiome}{S2}
\dummylabel{sec:supp_synthetic}{S3}

\end{document}